\documentclass[11pt]{article}
\usepackage{graphicx}
\usepackage{amsmath,amssymb,amsthm,amsxtra,overpic,bbm,bm,epsfig,ulem}
\usepackage{color}
\usepackage{multirow}
\usepackage{rotating}
\usepackage{slashbox}
\usepackage{subfigure}

\usepackage{array}
\newcommand{\PreserveBackslash}[1]{\let\temp=\\#1\let\\=\temp}
\newcolumntype{C}[1]{>{\PreserveBackslash\centering}p{#1}}
\newcolumntype{R}[1]{>{\PreserveBackslash\raggedleft}p{#1}}
\newcolumntype{L}[1]{>{\PreserveBackslash\raggedright}p{#1}}

\makeatletter

\newcommand{\Rmnum}[1]{\expandafter\@slowromancap\romannumeral #1@}
\makeatother

\begin{document}

\begin{center}
{\Large \bf Neutrino $\mu$-$\tau$ reflection symmetry and its breaking \\
in the minimal seesaw}
\end{center}

\vspace{0.05cm}

\begin{center}
{\bf Zhi-Cheng Liu, Chong-Xing Yue and Zhen-hua Zhao \footnote{E-mail: zhzhao@itp.ac.cn}  } \\
{Department of Physics, Liaoning Normal University, Dalian 116029, China}
\end{center}

\vspace{0.2cm}

\begin{abstract}
In this paper, we attempt to implement the neutrino $\mu$-$\tau$
reflection symmetry (which predicts $\theta^{}_{23} = \pi/4$ and $\delta = \pm \pi/2$
as well as trivial Majorana phases)
in the minimal seesaw (which enables us to fix the neutrino masses).
For some direct (the preliminary experimental hints towards
$\theta^{}_{23} \neq \pi/4$ and $\delta \neq - \pi/2$) and indirect
(inclusion of the renormalization group equation effect and implementation of
the leptogenesis mechanism) reasons, we particularly study the breakings of
this symmetry and their phenomenological consequences.
\end{abstract}

\newpage

\section{Introduction}

As is known, the discovery of neutrino oscillations indicates that neutrinos are
massive and mixed \cite{pdg}. On the one hand, the most popular way of
generating neutrino masses is to invoke the seesaw mechanism
which allows for a natural explanation of their smallness. In the type-\Rmnum{1}
seesaw mechanism \cite{seesaw}, one usually introduces three right-handed neutrino
fields $N^{}_{i}$ (for $i=1, 2, 3$).
They not only take part in Yukawa interactions with the left-handed neutrino fields
which lead to a Dirac mass matrix $M^{}_{\rm D}$, but also have a Majorana
mass matrix $M^{}_{\rm N}$ of themselves. Under the assumption of $M^{}_{\rm N} \gg
M^{}_{\rm D}$, the effective mass matrix for light neutrinos is obtained as
\cite{seesaw}
\begin{eqnarray}
M^{}_\nu = - M^{}_{\rm D} M^{-1}_{\rm N} M^{\rm T}_{\rm D} \;.
\label{1}
\end{eqnarray}
The overall minus sign here is of no physical meaning and will be neglected
in the following discussions. On the other hand, the neutrino mixing arises from a mismatch between
their mass and flavor eigenstates and is described by a $3 \times 3$ unitary matrix
$U= U^\dagger_l U^{}_\nu$ \cite{pmns} with $U^{}_l$ and $U^{}_\nu$ being respectively
the unitary matrix for diagonalizing the charged-lepton mass matrix $M^{}_l$ and $M^{}_\nu$.
In the standard parametrization, $U$ reads \cite{pdg}
\begin{eqnarray}
U \equiv P^{}_\phi V P^{}_\nu = P^{}_\phi
\left( \begin{matrix}
c^{}_{12} c^{}_{13} & s^{}_{12} c^{}_{13} & s^{}_{13} e^{-{\rm i} \delta} \cr
-s^{}_{12} c^{}_{23} - c^{}_{12} s^{}_{23} s^{}_{13} e^{{\rm i} \delta}
& c^{}_{12} c^{}_{23} - s^{}_{12} s^{}_{23} s^{}_{13} e^{{\rm i} \delta}  & s^{}_{23} c^{}_{13} \cr
s^{}_{12} s^{}_{23} - c^{}_{12} c^{}_{23} s^{}_{13} e^{{\rm i} \delta}
& -c^{}_{12} s^{}_{23} - s^{}_{12} c^{}_{23} s^{}_{13} e^{{\rm i} \delta} & c^{}_{23}c^{}_{13}
\end{matrix} \right) P^{}_\nu \;.
\label{2}
\end{eqnarray}
Here $P^{}_\phi = {\rm Diag}(e^{{\rm i} \phi^{}_1}, e^{{\rm i} \phi^{}_2}, e^{{\rm i} \phi^{}_3})$
consists of three unphysical phases that can be removed via the rephasing of charged-lepton
fields, while $P^{}_\nu = {\rm Diag}(e^{{\rm i} \rho}, e^{{\rm i} \sigma}, 1)$
contains two Majorana phases. Similar to the
CKM matrix, $V$ possesses a Dirac phase $\delta$ and three mixing angles
$\theta^{}_{ij}$ (for $ij=12, 13, 23$). For convenience, the abbreviations
$c^{}_{ij} = \cos{\theta^{}_{ij}}$ and $s^{}_{ij} = \sin{\theta^{}_{ij}}$ have been used.
Besides the mixing parameters, neutrino oscillations are also regulated by two
neutrino mass-squared differences $\Delta m^2_{ij} = m^2_i - m^2_j$ (for $ij = 21, 31$) \cite{global}
\begin{eqnarray}
\Delta m^2_{21}= (7.50 \pm 0.18) \times 10^{-5} \ {\rm eV}^2 \;, \hspace{1cm}
|\Delta m^2_{31}| = (2.524 \pm 0.039) \times 10^{-3} \ {\rm eV}^2 \;.
\label{3}
\end{eqnarray}
In comparison, the absolute neutrino mass scale has to be inferred from non-oscillatory
experiments \cite{non-osci}, and a definite result is still missing.
Note that the sign of $\Delta m^2_{31}$ has not yet been determined, thereby allowing
for two possible mass orderings $m^{}_1 < m^{}_2 < m^{}_3$ (the normal hierarchy, NH
for short) and $m^{}_3 < m^{}_1 < m^{}_2$ (the inverted hierarchy, IH for short).
It turns out that the fitted values of mixing parameters depends on the mass
ordering in a certain way: $\theta^{}_{13}$, $\theta^{}_{23}$ and $\delta$ take the values
\begin{eqnarray}
\sin^2{\theta^{}_{13}} = 0.02166 \pm 0.00075  \;, \hspace{1cm}
\sin^2{\theta^{}_{23}} = 0.441 \pm 0.024 \;, \hspace{1cm}
\delta = 261^\circ \pm 55^\circ \;,
\label{4}
\end{eqnarray}
in the NH case, or
\begin{eqnarray}
\sin^2{\theta^{}_{13}} = 0.02179 \pm 0.00076  \;, \hspace{1cm}
\sin^2{\theta^{}_{23}} = 0.587 \pm 0.022 \;, \hspace{1cm}
\delta = 277^\circ \pm 43^\circ \;,
\label{5}
\end{eqnarray}
in the IH case, whereas $\theta^{}_{12}$ takes the value $\sin^2{\theta^{}_{12}} = 0.306 \pm 0.012$
in either case \cite{global}.

How to understand the observed neutrino mixing poses an interesting question.
As one can see, the measured $\theta^{}_{12}$, $\theta^{}_{23}$ and $\delta$
are close to some special values
\begin{eqnarray}
\sin^2{\theta^{}_{12}} = \frac{1}{3}  \;, \hspace{1cm}
\sin^2{\theta^{}_{23}} = \frac{1}{2}  \;, \hspace{1cm}
\delta= - \frac{\pi}{2} \;.
\label{6}
\end{eqnarray}
These remarkable coincidences invite
us to speculate that some flavor symmetry has played an important role in
shaping the neutrino mixing \cite{review}.
In this connection, the $\mu$-$\tau$ reflection symmetry \cite{MTR} serves as a
unique candidate: In the basis of $M^{}_l$ being diagonal,
$M^{}_\nu$ should stay invariant with respect to the transformations
\footnote{This operation is a combination of the $\mu$-$\tau$ exchange and CP conjugate
transformations --- a specific example of the generalized CP transformations \cite{gcp}.}
\begin{eqnarray}
\nu^{}_e \leftrightarrow \nu^c_e \;, \hspace{1cm} \nu^{}_\mu \leftrightarrow \nu^c_\tau \;,
\hspace{1cm} \nu^{}_\tau \leftrightarrow \nu^c_\mu \;,
\label{7}
\end{eqnarray}
and have its elements $M_{\alpha\beta}$ (for $\alpha, \beta
= e, \mu, \tau$) obeying the conditions
\begin{eqnarray}
M^{}_{e\mu} = M^*_{e\tau} \;, \hspace{1cm} M^{}_{\mu\mu} = M^*_{\tau\tau}  \;, \hspace{1cm}
M^{}_{ee} \ {\rm and} \ M^{}_{\mu\tau} \ {\rm being \ real}  \;.
\label{8}
\end{eqnarray}
Consequently, $U^{}_l$ is a diagonal phase matrix of no physical meaning
$U^{}_l = {\rm Diag} (e^{{\rm i} \varphi^{}_1}, e^{{\rm i} \varphi^{}_2},
e^{{\rm i} \varphi^{}_3})$ where $\varphi^{}_{1, 2, 3}$ can take arbitrary values.
On the other hand, as a result of the six real conditions for $M^{}_\nu$
given by Eq. (1), there are six predictions \cite{GL}
\begin{eqnarray}
\phi^{}_{1} = 0 \;, \hspace{1cm} \phi^{}_{ 2} = - \phi^{}_{ 3} \;,
\hspace{1cm} \theta^{}_{ 23} = \pi/4 \;, \hspace{1cm}
\delta^{} = \pm \pi/2 \;, \hspace{1cm} \rho, \sigma = 0 \ {\rm or} \ \pi/2 \;,
\label{9}
\end{eqnarray}
for the mixing parameters of $U^{}_\nu$ which itself takes a form as given by Eq. (\ref{2}).
Furthermore, unlike the $\mu$-$\tau$ interchange symmetry \cite{MT,review2}
which predicts $\theta^{}_{13} =0$, the $\mu$-$\tau$ reflection symmetry
allows for an arbitrary $\theta^{}_{13}$.
Thanks to these interesting consequences, this symmetry
has been attracting a lot of attention recently \cite{MTRs}.

Nevertheless, the neutrino masses receive no clues from the $\mu$-$\tau$ reflection
symmetry. But they will be fixed if the smallest one ($m^{}_1$ in the NH
case, $m^{}_3$ in the IH case) is to vanish, a possibility that is experimentally
allowed and can be accommodated by the minimal seesaw
\cite{mss} where only two right-handed neutrino fields $N^{}_{1, 2}$ are involved.
In this paper, we just attempt to combine the $\mu$-$\tau$ reflection
symmetry with the minimal seesaw so that both the neutrino mixing and mass
spectrum can be settled. The implications of such a marriage
are discussed in the next section. In consideration of the preliminary
experimental results towards $\theta^{}_{23} \neq \pi/4$
\footnote{In particular, a recent result reported by the NOvA collaboration
($\theta^{}_{23} = 39.5^\circ \pm 1.7^\circ$ or $52.1^\circ \pm 1.7^\circ$ in the NH case)
disfavors the possibility of $\theta^{}_{23}= 45^\circ$ at a 2.6$\sigma$
level \cite{NOvA}.} and $\delta \neq - \pi/2$, we particularly explore the
possible symmetry breakings and their impacts on the mixing parameters
in section 3. Section 4 is devoted to the specific symmetry breaking triggered
by the renormalization group equation (RGE) effect. In section 5, the operation
of leptogenesis in this framework will be studied. Finally,
our main results are summarized in section 6.

\section{$\mu$-$\tau$ reflection symmetry in the minimal seesaw}

Conversely, one can reconstruct an $M^{}_\nu$ of the $\mu$-$\tau$ reflection symmetry
in terms of the $U^{}_\nu$ characterized by Eq. (2) and neutrino masses
by means of the relation
$M^{}_\nu = U^{}_\nu {\rm Diag}(m^{}_1, m^{}_2, m^{}_3) U^{\rm T}_\nu$.
In the situation of one neutrino mass being vanishing, the reconstructed $M^{}_\nu$
in such a way will take a form as
\begin{eqnarray}
M^{}_{ee} & = &  \overline m^{}_2 s^2_{12} c^2_{13} - m^{}_3 s^{2}_{13} \; ,
\nonumber \\
M^{}_{e\mu} & = & \frac{1}{\sqrt 2} \left[ \overline m^{}_2 s^{}_{12}  \left( c^{}_{12} + {\rm i}
s^{}_{12} \bar s^{}_{13} \right) + {\rm i}  m^{}_3 \bar s^{}_{13} \right] c^{}_{13} e^{{\rm i}\phi} \;,
\nonumber \\
M^{}_{\mu\mu} & = & \frac{1}{2} \left[ \overline m^{}_2 \left( c^{}_{12} + {\rm i} s^{}_{12}
\bar s^{}_{13} \right)^2 + m^{}_3 c^2_{13} \right] e^{2{\rm i}\phi} \;,
\nonumber \\
M^{}_{\mu\tau} & = & \frac{1}{2} \left[ \overline m_2^{} \left( c^{2}_{12} +
s^{2}_{12} s^{2}_{13} \right) - m^{}_3 c^2_{13} \right] \;,
\label{11}
\end{eqnarray}
in the NH case, or
\begin{eqnarray}
M^{}_{ee} & = & \left[ m^{}_1 c^2_{12} + \overline m^{}_2 s^2_{12}  \right] c^2_{13}  \; ,
\nonumber \\
M^{}_{e\mu} & = & \frac{1}{\sqrt 2} \left[m^{}_1 c^{}_{12} \left( -s^{}_{12}
+ {\rm i}  c^{}_{12} \overline s^{}_{13} \right)
+ \overline m^{}_2 s^{}_{12} \left( c^{}_{12}
+ {\rm i}  s^{}_{12} \overline s^{}_{13}  \right) \right] c^{}_{13} e^{ {\rm i} \phi} \;,
\nonumber \\
M^{}_{\mu\mu} & = & \frac{1}{2} \left[  m^{}_1 \left( s^{}_{12}
- {\rm i} c^{}_{12} \overline s^{}_{13}  \right)^2 + \overline m^{}_2 \left(  c^{}_{12}
+ {\rm i} s^{}_{12} \overline s^{}_{13} \right)^2 \right]  e^{2 {\rm i}  \phi}  \; ,
\nonumber \\
M^{}_{\mu\tau} & = & \frac{1}{2} \left[ m^{}_1 \left( s^2_{12} + c^2_{12} s^2_{13} \right)
+ \overline m^{}_2 \left( c^2_{12} + s^2_{12} s^2_{13}  \right) \right]  \;,
\label{12}
\end{eqnarray}
in the IH case, where $\overline{m}^{}_2 \equiv m^{}_2 e^{2{\rm i} \sigma}$
and $\bar {s}^{}_{13} \equiv -{\rm i} s^{}_{13} e^{{\rm i} \delta}$ have been defined.
Note that $M^{}_{e\mu}$ and $M^{}_{\mu \mu}$ depend on the unphysical phase
$\phi \equiv \phi^{}_2 = - \phi^{}_3$ whose value can be chosen freely
without affecting the physical results. And there is only one effective Majorana phase
which we assign for $m^{}_2$ (i.e., $\sigma$).
Because of the condition ${\rm Det}(M^{}_\nu) =0$
\footnote{Since the determinant of an $M^{}_\nu$ obeying the $\mu$-$\tau$ reflection
symmetry is always real, this condition only amounts to one constraint.},
only five out of the six real components of these elements are independent. Taking
the best-fit values for $\theta^{}_{13}$,
$\theta^{}_{12}$, $\Delta m^2_{21}$ and $|\Delta m^2_{31}|$ as input, we give
the values of $M^{}_{\alpha\beta}$ for various combinations of
$\delta$ and $\sigma$ (i.e., $[\delta, \sigma] = [\pi/2, 0], [-\pi/2, 0],
[\pi/2, \pi/2]$ and $[-\pi/2, \pi/2]$) in the NH (Table 1) and IH (Table 2) cases.
As is well known, the size of $|M^{}_{ee}|$ which governs the rate of neutrino-less
double beta decays \cite{0nbb} is much larger in the IH case than in the NH case.
In the NH case, the elements exhibit a hierarchical structure as
$|M^{}_{\mu \mu}| \sim |M^{}_{\mu \tau}| \gg |M^{}_{e e}| \sim |M^{}_{e\mu }|$,
implying that they might have received contributions from different levels
\cite{mohapatra1}. In the IH case, $|M^{}_{ee}|$ (so does $|M^{}_{e\mu}|$ for
$\sigma = 0$) becomes comparable to $|M^{}_{\mu \mu}|$ and $|M^{}_{\mu \tau}|$.
But for $\sigma = \pi/2$, $M^{}_{e\mu}$ will have a magnitude much smaller than
the other ones as a result of the heavy cancellation between its two components
respectively associated with $m^{}_1$ and $\overline m^{}_2$.
By choosing the value of $\phi$ in such a way that the phase of $M^{}_{e\mu}$
or $M^{}_{\mu \mu}$ cancels out, one may convert $M^{}_\nu$
to a simpler form as given by Tables 3-4.

\begin{table}[h]
\centering
\begin{tabular}{|p{3.5cm}<{\centering}|p{2.cm}<{\centering}|p{3.cm}<{\centering}
|p{3.cm}<{\centering}|p{2.cm}<{\centering}|} \hline
\backslashbox{[$\delta$, $\sigma$]} {$M^{}_{\alpha\beta}$(eV)} & $M^{}_{ee}(\times 10^{-2})$
& $M^{}_{e\mu}(\times 10^{-2}e^{ {\rm i} \phi})$ & $M^{}_{\mu\mu}(\times 10^{-2}e^{2 {\rm i} \phi})$
&  $M^{}_{\mu\tau}(\times 10^{-2})$ \\ \hline \hline
$[\pi/2, 0]    $ & $ 0.15 $ & $ 0.28 + 0.54 {\rm i} $ & $ 2.76 + 0.06 {\rm i} $ & $ -2.15 $  \\ \hline
$[-\pi/2, 0]   $ & $ 0.15 $ & $ 0.28 - 0.54 {\rm i} $ & $ 2.76 - 0.06 {\rm i} $ & $ -2.15 $  \\ \hline
$[\pi/2,\pi/2] $ & $-0.37 $ & $-0.28 + 0.49 {\rm i} $ & $ 2.16 - 0.06 {\rm i} $ & $ -2.76 $   \\ \hline		
$[-\pi/2,\pi/2]$ & $-0.37 $ & $-0.28 - 0.49 {\rm i} $ & $ 2.16 + 0.06 {\rm i} $ & $ -2.76 $ \\ \hline
\end{tabular}
\caption{The possible values of $M^{}_{\alpha\beta}$ for various combinations of
$\delta$ and $\sigma$ in the NH case.}
\end{table}

\begin{table}[h]
\centering
\begin{tabular}{|p{3.5cm}<{\centering}|p{2.cm}<{\centering}|p{3.cm}<{\centering}
|p{3.cm}<{\centering}|p{2.cm}<{\centering}|} \hline
\backslashbox{[$\delta$, $\sigma$]}{$M^{}_{\alpha \beta}$(eV)} & $M^{}_{ee}(\times 10^{-2})$
& $M^{}_{e\mu}(\times 10^{-2}e^{ {\rm i} \phi})$ & $M^{}_{\mu\mu}(\times 10^{-2}e^{2 {\rm i} \phi})$
&  $M^{}_{\mu\tau}(\times 10^{-2})$ \\ \hline \hline
$[\pi/2, 0]    $ & $ 4.86 $ & $ 0.02 + 0.51 {\rm i} $ & $ 2.45 + 0.01 {\rm i} $ & $ 2.56 $  \\ \hline
$[-\pi/2, 0]   $ & $ 4.86 $ & $ 0.02 - 0.51 {\rm i} $ & $ 2.45 - 0.01 {\rm i} $ & $ 2.56 $  \\ \hline
$[\pi/2,\pi/2] $ & $ 1.86 $ & $-3.21 + 0.20 {\rm i} $ & $-1.01 - 0.68 {\rm i} $ & $-0.97 $  \\ \hline
$[-\pi/2,\pi/2]$ & $ 1.86 $ & $-3.21 - 0.20 {\rm i} $ & $-1.01 + 0.68 {\rm i} $ & $-0.97 $ \\ \hline
\end{tabular}
\caption{The possible values of $M^{}_{\alpha\beta}$ for various combinations of
$\delta$ and $\sigma$ in the IH case.}
\end{table}

\begin{table}[h]
\centering
\begin{tabular}{|p{3.5cm}<{\centering}||p{2.cm}<{\centering}|p{3.cm}<{\centering}
||p{3.cm}<{\centering}|p{2.cm}<{\centering}|} \hline
\backslashbox{[$\delta$, $\sigma$]} {$M^{}_{\alpha \beta}$(eV)} & $M^{}_{e\mu}(\times 10^{-2})$
& $M^{}_{\mu\mu}(\times 10^{-2})$ & $M^{}_{e\mu}(\times 10^{-2})$ &  $M^{}_{\mu\mu}(\times 10^{-2})$ \\ \hline \hline
$[\pi/2, 0]$     & $ 0.61 $ & $ -1.56 - 2.27 {\rm i} $ & $ 0.28 + 0.54 {\rm i} $ & $ 2.76 $  \\ \hline
$[-\pi/2, 0]$    & $ 0.61 $ & $ -1.56 + 2.27 {\rm i} $ & $ 0.28 - 0.54 {\rm i} $ & $ 2.76 $  \\ \hline
$[\pi/2, \pi/2]$ & $ 0.56 $ & $ -1.05 + 1.89 {\rm i} $ & $-0.29 + 0.49 {\rm i} $ & $ 2.16 $   \\ \hline
$[-\pi/2,\pi/2]$ & $ 0.56 $ & $ -1.05 - 1.89 {\rm i} $ & $-0.29 - 0.49 {\rm i} $ & $ 2.16 $ \\ \hline 	
\end{tabular}
\caption{The possible values of $M^{}_{e\mu}$ and $M^{}_{\mu\mu}$ for various combinations
of $\delta$ and $\sigma$ in the NH case, after one of them is made real by a particular
value of $\phi$. }
\end{table}

\begin{table}[h]
\centering
\begin{tabular}{|p{3.5cm}<{\centering}||p{2.cm}<{\centering}|p{3.cm}<{\centering}
||p{3.cm}<{\centering}|p{2.cm}<{\centering}|} \hline
\backslashbox{[$\delta$, $\sigma$]} {$M^{}_{\alpha \beta}$(eV)} & $M^{}_{e\mu}(\times 10^{-2})$
& $M^{}_{\mu\mu}(\times 10^{-2})$ & $M^{}_{e\mu}(\times 10^{-2})$ &  $M^{}_{\mu\mu}(\times 10^{-2})$ \\ \hline \hline
$[\pi/2, 0]$     & $ 0.51 $ & $ -2.43 - 0.24 {\rm i} $ & $ 0.02 + 0.51 {\rm i} $ & $ 2.45 $  \\ \hline
$[-\pi/2, 0]$    & $ 0.51 $ & $ -2.43 + 0.24 {\rm i} $ & $ 0.02 - 0.51 {\rm i} $ & $ 2.45 $  \\ \hline
$[\pi/2, \pi/2]$ & $ 3.22 $ & $ -0.92 - 0.80 {\rm i} $ & $-1.13 - 3.02 {\rm i} $ & $ 1.21 $   \\ \hline
$[-\pi/2,\pi/2]$ & $ 3.22 $ & $ -0.92 + 0.80 {\rm i} $ & $-1.13 + 3.02 {\rm i} $ & $ 1.21 $ \\ \hline 		
\end{tabular}
\caption{The possible values of $M^{}_{e\mu}$ and $M^{}_{\mu\mu}$ for various combinations
of $\delta$ and $\sigma$ in the IH case, after one of them is made real by a particular
value of $\phi$. }
\end{table}

The $M^{}_\nu$ given by Eqs. (\ref{11}-\ref{12}) can be viewed as a result of the
minimal seesaw:
The Dirac mass matrix coupling $N^{}_{1, 2}$ with the left-handed neutrino fields
is assumed to appear as \cite{minimal}
\begin{eqnarray}
M^{}_{\rm D} = \left( \begin{matrix}
& a^{}_1 \sqrt{M^{}_1}   \hspace{0.4cm} \sqrt{M^{}_2} b^{}_1  \cr
& e^{{\rm i} \phi^{}_a}  a^{}_2 \sqrt{M^{}_1}   \hspace{0.4cm} \sqrt{M^{}_2} b^{}_2 e^{{\rm i} \phi^{}_b}  \cr
& e^{-{\rm i} \phi^{}_a} a^{}_2 \sqrt{M^{}_1}  \hspace{0.4cm} \sqrt{M^{}_2} b^{}_2 e^{-{\rm i} \phi^{}_b}
\end{matrix} \right) \;,
\label{13}
\end{eqnarray}
with $a^{}_{1, 2}$, $b^{}_{1, 2}$, $\phi^{}_{a, b}$ and $M^{}_{1, 2}$ being real parameters.
Apparently, its elements satisfy the conditions of $M^{}_{\mu i} = M^{*}_{\tau i}$ and
$M^{}_{ei}$ being real (for $i=1, 2$). It is easy to see that these
conditions still hold when the right-handed neutrino fields experience an orthogonal
basis transformation. So, without loss of generality, we choose to work in the basis
where the Majorana mass matrix for $N^{}_{1, 2}$ is diagonal
$M^{}_{\rm N} = {\rm Diag} (M^{}_1, M^{}_2)$. By virtue of the seesaw formula in
Eq. (\ref{1}), we arrive at an effective neutrino mass matrix
\begin{eqnarray}
M^{}_\nu = \left( \begin{matrix}
\vspace{0.2cm}
a^2_1 + b^2_1 & \hspace{0.2cm} a^{}_1 a^{}_2  e^{{\rm i} \phi^{}_a}  + b^{}_1 b^{}_2  e^{{\rm i} \phi^{}_b}
& \hspace{0.2cm} a^{}_1 a^{}_2  e^{-{\rm i} \phi^{}_a} + b^{}_1 b^{}_2 e^{-{\rm i} \phi^{}_b}   \cr
\vspace{0.2cm}
a^{}_1 a^{}_2 e^{{\rm i} \phi^{}_a} + b^{}_1 b^{}_2 e^{{\rm i} \phi^{}_b} & \hspace{0.2cm}
a^{2}_2 e^{2 {\rm i} \phi^{}_a} + b^{2}_2 e^{2{\rm i} \phi^{}_b}
& \hspace{0.2cm} a^{2}_2 + b^{2}_2 \cr
\vspace{0.2cm}
a^{}_1 a^{}_2 e^{- {\rm i} \phi^{}_a}  + b^{}_1 b^{}_2 e^{- {\rm i} \phi^{}_b}  & \hspace{0.2cm} a^{2}_2 + b^{2}_2
& \hspace{0.2cm} a^{2}_2 e^{-2 {\rm i} \phi^{}_a} + b^{2}_2 e^{-2{\rm i} \phi^{}_b}
\end{matrix} \right) \;.
\label{14}
\end{eqnarray}
Diagonalizing this $M^{}_\nu$
with a $U^{}_\nu$ characterized by Eq. (\ref{9}) yields the mixing parameters
\begin{eqnarray}
\tan \theta^{}_{13} & = & \frac{ a^2_2 \sin \left( 2 \phi - 2 \phi^{}_a \right)
+ b^2_2 \sin \left( 2 \phi - 2 \phi^{}_b \right) } {- \sqrt{2} \sin \delta
\left[ a^{}_1 a^{}_2 \cos \left( \phi - \phi^{}_a \right)
+ b^{}_1 b^{}_2 \cos \left( \phi - \phi^{}_b \right) \right]} \; ,  \nonumber \\
\tan 2 \theta_{13} & = & \frac{-2 \sqrt{2} \left[ a^{}_1 a^{}_2 \sin \left( \phi - \phi^{}_a \right)
+ b^{}_1 b^{}_2 \sin \left( \phi - \phi^{}_b \right) \right] } {\sin \delta \left[  a^2_1 + b^2_1 - P \right]} \; ,
\nonumber \\
\tan 2 \theta^{}_{12} & = &  \frac { -2 \sqrt {2}  \cos 2 \theta^{}_{13}
\left[ a^{}_1 a^{}_2 \cos \left( \phi - \phi^{}_a \right) + b^{}_1 b^{}_2 \cos
\left( \phi - \phi^{}_b \right) \right] } { c^{}_{13} \left[ \left( a^2_1 +b^2_1 \right) c^2_{13}
- P  s^2_{13} - Q  \cos 2 \theta^{}_{13} \right] }\; ,
\label{15a}
\end{eqnarray}
and neutrino masses
\begin{eqnarray}
\overline m^{}_1  &=& Q - \frac { \sqrt 2} { c^{}_{13} t^{}_{12} }
\left[ a^{}_1 a^{}_2 \cos ( \phi - \phi^{}_a  )  + b^{}_1 b^{}_2 \cos ( \phi - \phi^{}_b  ) \right] \; ,
\nonumber \\
\overline m^{}_2 &=& Q + \frac { \sqrt 2 t^{}_{12}  } { c^{}_{13} }
\left[  a^{}_1 a^{}_2 \cos ( \phi - \phi^{}_a  )  + b^{}_1 b^{}_2 \cos ( \phi - \phi^{}_b  ) \right] \; ,
\nonumber \\
\overline m^{}_3 &=& \frac { \left( a^2_1 + b^2_1  \right) s^2_{13} - P  c^2_{13} }{\cos 2 \theta^{}_{13}}\; ,
\label{15b}
\end{eqnarray}
where $\overline m^{}_1  =  m^{}_1 e^{2{\rm i}\rho}$, $\overline m^{}_3  =  m^{}_3 e^{2{\rm i} \gamma}$
(for $\rho, \gamma= 0$ or $\pi/2$) and
\begin{eqnarray}
P & = &  2 a^2_2 \sin^2 \left( \phi - \phi^{}_a \right) + 2 b^2_2 \sin^2 \left( \phi - \phi^{}_b   \right)\; ,
\nonumber\\
Q & = & 2 a^2_2 \cos^2 \left( \phi - \phi^{}_a \right) + 2 b^2_2 \cos^2 \left( \phi - \phi^{}_b   \right)\; .
\label{16}
\end{eqnarray}
For any given values of $a^{}_{1, 2}$, $b^{}_{1, 2}$ and $\phi^{}_{a, b}$, one mass
will necessarily vanish as promised by the minimal seesaw
\footnote{ We are left with the difference of the phases associated with two non-zero
masses as the effective Majorana phase.}.
The resulting $\theta^{}_{13}$, $\phi$, $\theta^{}_{12}$
and two non-zero masses can be calculated with the help of the other five equations.

If we are to derive the allowed values of $a^{}_{1, 2}$, $b^{}_{1, 2}$
and $\phi^{}_{a, b}$ from the measured $\theta^{}_{13}$, $\theta^{}_{23}$,
$\Delta m^2_{21}$ and $|\Delta m^2_{31}|$, one just needs to confront the
$M^{}_\nu$ in Eq. (\ref{14}) with the results given by Tables 1-2.
Above all, it should be noted that both the $ee$ and $\mu \tau$ elements of
this $M^{}_\nu$ are positive. In order for them to fit in with
the corresponding results in Tables 1-2, one must have $\sigma = \pi/2$ (or 0) and
$M^{}_\nu \to -M^{}_\nu \ ({\rm or} \ +M^{}_\nu)$ in the NH (or IH) case.
In light of the unphysical nature of $\phi$, $\phi^{}_{a, b} - \phi$ rather than
$\phi^{}_{a, b}$ will be treated as effective independent parameters.
Recall that only five real components
of the neutrino mass matrix elements are independent, so the free parameters
are more than the constraint equations by one.
For this reason, in Fig. 1 we choose to present the results for $a^{}_{2}$, $b^{}_{1, 2}$
and $\phi^{}_{a, b} -\phi$ as functions of $a^{}_1$ (which
stands in an equivalent position as $b^{}_1$). In the numerical calculations
here and in the following, the best-fit values for $\theta^{}_{13}$, $\theta^{}_{12}$,
$\Delta m^2_{21}$ and $|\Delta m^2_{31}|$ are input, whereas
$\delta$ is specified as $-\pi/2$.
We have only shown the results in the case of both $a^{}_{1}$ and $b^{}_1$ being positive.
The results in the case of $a^{}_{1}$ or (and) $b^{}_1$ being negative can be
obtained by simply making the replacement $a^{}_1 \to - a^{}_1$ combined with
$(\phi^{}_a- \phi) \to (\phi^{}_a- \phi) + \pi$ or (and) $b^{}_1 \to - b^{}_1$ combined with
$(\phi^{}_b- \phi) \to (\phi^{}_b - \phi)+ \pi$. This is because the $M^{}_\nu$ in Eq. (\ref{14})
keeps invariant under this kind of transformations.
It is interesting to find that the possibility of $a^{}_1 =0$ is allowed.
The possible values of $a^{}_2$, $b^{}_{1, 2}$ and $\phi^{}_{a, b}- \phi$ in such a
particular case are listed in Table 5. If we further make one of $\phi^{}_{a, b}$ vanish
by giving $\phi$ an appropriate value, then we will reach the simplest $M^{}_{\rm D}$.
In the NH case, for instance, a value of $-1.79$ or $-4.93$ for $\phi$ allows
us to have $\phi^{}_{a} = 0$. (In the meantime, $\phi^{}_{b}$ is fixed to $-0.74$ or $-3.88$.)

\begin{figure}
\centering
\includegraphics[width=6in]{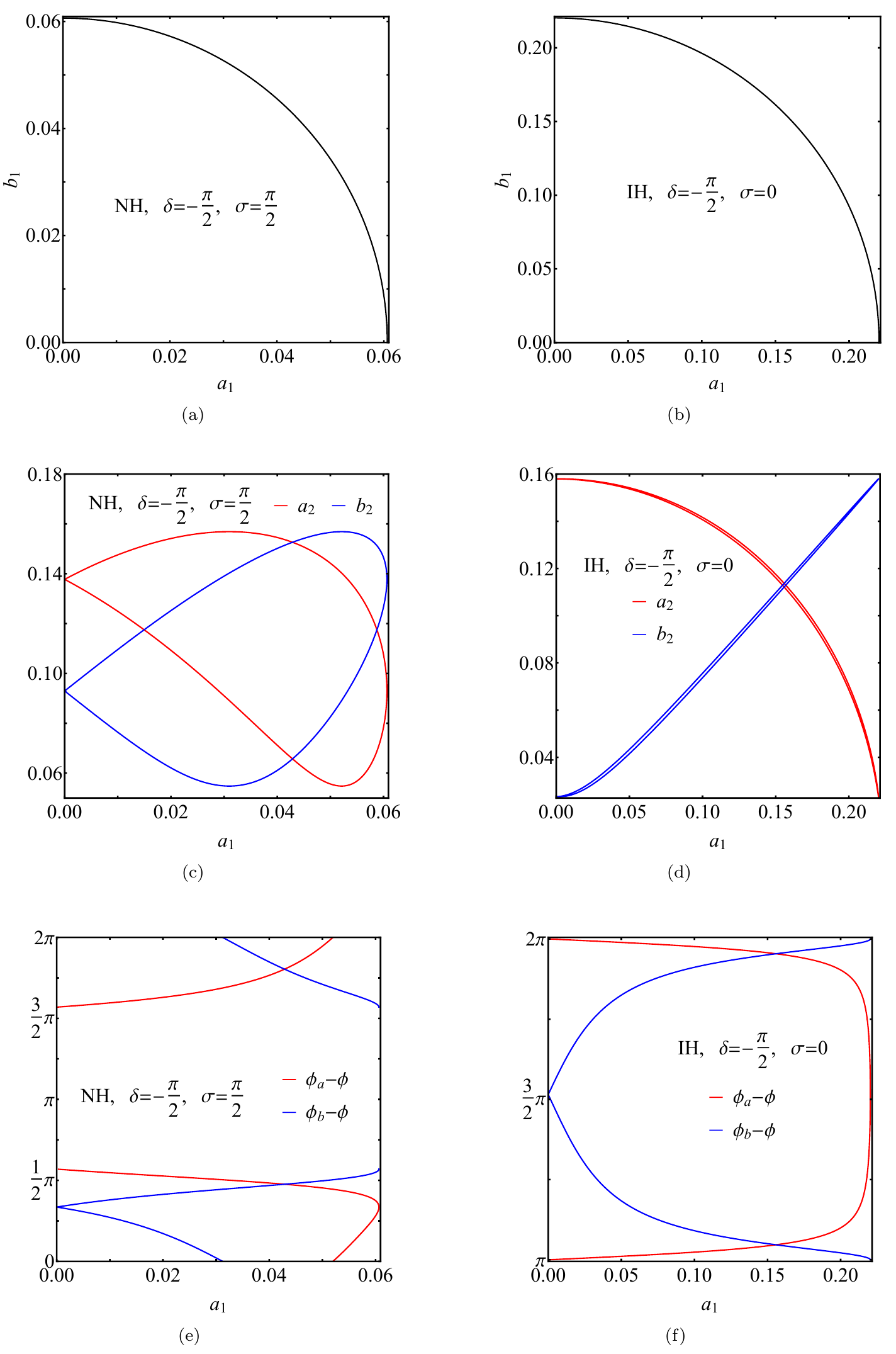}
\caption{ The allowed values of $a^{}_2$, $b^{}_{1, 2}$ and $\phi^{}_{a, b} - \phi$
as functions of $a^{}_1$, with $\sigma = \pi/2$ (or 0) in the NH (or IH) case and
$\delta = -\pi/2$. }
\label{fig:subfig} 
\end{figure}

\begin{table}[h]
\centering
\begin{tabular}{|p{1.5cm}<{\centering}|p{1.5cm}<{\centering}|p{1.5cm}<{\centering}
|p{1.5cm}<{\centering}|p{1.5cm}<{\centering}|p{3.cm}<{\centering}|} \hline
$  $ & $b^{}_1$ & $\phi^{}_b - \phi$ & $a^{}_2$ &  $b^{}_2$ &  $\phi^{}_a - \phi $\\ \hline \hline
NH & \  $0.06 $&$1.05 $&$ 0.14 $&$ 0.09$&$ 1.79,4.93$ \\ \hline
IH &\ $0.22 $&$ 4.76 $&$ 0.16 $&$ 0.02$&$ 0, \pi$ \\ \hline				
\end{tabular}
\caption{The possible values of $a^{}_2$, $b^{}_{1, 2}$ and $\phi^{}_{a, b}- \phi$
in the particular case of $a^{}_1 =0$.}
\end{table}

\section{Breaking of the $\mu$-$\tau$ reflection symmetry}

In this section we study the possible breakings of $\mu$-$\tau$ reflection symmetry
and their impacts on the mixing parameters \cite{GJKLST}.
The most general perturbation to $M^{}_{\rm D}$
\begin{eqnarray}
\delta M^{}_{\rm D} = \left( \begin{matrix}
\delta^{}_{11} \sqrt{M^{}_1}  & \hspace{0.4cm} \sqrt{M^{}_2} \delta^{}_{12}  \cr
\delta^{}_{21} \sqrt{M^{}_1}  & \hspace{0.4cm} \sqrt{M^{}_2} \delta^{}_{22}  \cr
\delta^{}_{31} \sqrt{M^{}_1}  & \hspace{0.4cm} \sqrt{M^{}_2} \delta^{}_{32}
\end{matrix} \right) \;,
\label{17}
\end{eqnarray}
can be decomposed into a symmetry-conserving part and a symmetry-violating part:
\begin{eqnarray}
\delta M^{}_{\rm D} =\frac {1} {2} \left( \begin{matrix}
2 {\rm Re} (\delta^{}_{11} ) \sqrt{M^{}_1} & 2 {\rm Re}( \delta^{}_{12} ) \sqrt{M^{}_2}  \\
(\delta^{}_{21} + \delta_{31}^*) \sqrt{M^{}_1} & (\delta^{}_{22}+\delta_{32}^*) \sqrt{M^{}_2}\\
(\delta_{21}^* + \delta^{}_{31}) \sqrt{M^{}_1} & (\delta_{22}^*+\delta^{}_{32})\sqrt {M^{}_2}
\end{matrix} \right)
+\frac {1} {2} \left( \begin{matrix}
2 {\rm i Im} ( \delta^{}_{11} ) \sqrt{M^{}_1}  & 2 {\rm i Im}( \delta^{}_{12} ) \sqrt{M^{}_2}  \\
( \delta^{}_{21} - \delta_{31}^* ) \sqrt{M^{}_1} & ( \delta^{}_{22} - \delta_{32}^* ) \sqrt{M^{}_2}\\
( \delta^{}_{31} - \delta_{21}^* ) \sqrt{M^{}_1} & ( \delta^{}_{32} - \delta_{22}^* ) \sqrt {M^{}_2}
\end{matrix} \right) \;.
\label{18}
\end{eqnarray}
And the full Dirac mass matrix can be reparameterized as
\begin{eqnarray}
M^{\prime}_{\rm D} = M^{}_{\rm D} + \delta M^{}_{\rm D}
=\left( \begin{matrix}
\hat a^{}_1 ( 1 + {\rm i} \epsilon^{}_1 )  \sqrt{M^{}_1}
& \hspace{0.4cm}  \hat b^{}_1  ( 1 + {\rm i} \epsilon^{}_2 )  \sqrt{M^{}_2}  \cr
\hat a^{}_2  e^{{\rm i} \hat \phi^{}_a}  ( 1 + \epsilon^{}_3 )  \sqrt{M^{}_1}
& \hspace{0.4cm} \hat b^{}_2  e^{{\rm i} \hat \phi^{}_b}  ( 1 + \epsilon^{}_4 ) \sqrt{M^{}_2}  \cr
\hat a^{}_2  e^{ - {\rm i} \hat \phi^{}_a}  ( 1 - \epsilon_3^* )  \sqrt{M^{}_1} &
\hspace{0.4cm} \hat b^{}_2  e^{ - {\rm i} \hat \phi^{}_b}  ( 1 - \epsilon_4^* )  \sqrt{M^{}_2}
\end{matrix} \right) \;,
\label{19}
\end{eqnarray}
with
\begin{eqnarray}
& \hat a^{}_1 = a^{}_1 + {\rm Re}( \delta^{}_{11} ), & \hspace{1cm}
\hat a^{}_2  e^{{\rm i} \hat \phi^{}_a} = a^{}_2  e^{{\rm i} \phi^{}_a}
+ \frac { \delta^{}_{21} + \delta_{31}^* } {2} \;, \nonumber \\
& \hat b^{}_1 = b^{}_1 + {{\rm Re}}( \delta^{}_{21} ), & \hspace{1cm}
\hat b^{}_2  e^{{\rm i} \hat \phi^{}_b} = b^{}_2  e^{{\rm i} \phi^{}_b}
+ \frac { \delta^{}_{22} + \delta_{32}^* } {2} \;,
\label{20}
\end{eqnarray}
and
\begin{eqnarray}
&\epsilon^{}_1 \equiv \displaystyle \frac { {\rm Im} ( (M^{}_{\rm D})^{}_{e1} ) }
{ {\rm Re} ( (M^{}_{\rm D})^{}_{e1} )}
= \frac { {\rm Im}( \delta^{}_{11} ) }{ a^{}_1 + {\rm Re}(\delta^{}_{11}) } \;, \hspace{1cm}
\epsilon^{}_3 \equiv \frac{ (M^{}_{\rm D})^{}_{\mu1} -  (M^{}_{\rm D})^{*}_{\tau1}}
{(M^{}_{\rm D})^{}_{\mu1} +  (M^{}_{\rm D})^{*}_{\tau1}}
= \frac{ \delta^{}_{21} - \delta_{31}^*} { 2 a^{}_2 e^{ {\rm i} \phi^{}_a }
+ \delta^{}_{21} + \delta_{31}^* } \;, \nonumber \\
& \epsilon^{}_2 \equiv \displaystyle \frac { {\rm Im} ( (M^{}_{\rm D})^{}_{e2} ) }
{ {\rm Re} ( (M^{}_{\rm D})^{}_{e2} )}
= \frac { {\rm Im}( \delta^{}_{21} ) }{ a^{}_1 + {\rm Re} (\delta^{}_{21}) }  \;, \hspace{1cm}
\epsilon^{}_4 \equiv \frac{ (M^{}_{\rm D})^{}_{\mu 2} -  (M^{}_{\rm D})^*_{\tau 2}}
{(M^{}_{\rm D})^{}_{\mu 2} +  (M^{}_{\rm D})^*_{\tau2}}
= \frac{ \delta^{}_{22} - \delta_{32}^*} { 2 b^{}_2 e^{ {\rm i} \phi^{}_b }
+ \delta^{}_{22} + \delta_{32}^* }  \;.
\label{21}
\end{eqnarray}
The dimensionless quantities $\epsilon^{}_{1, 2, 3, 4}$ measure the strength of symmetry breaking.
They should be small (e.g., $ | \epsilon_{1, 2, 3, 4} | \leq 0.1 $)
in order for $M^\prime_{\rm D}$ to assume an approximate $\mu$-$\tau$ reflection symmetry.

Since $\epsilon^{}_2$ and $\epsilon^{}_4$ play equivalent roles as $\epsilon^{}_1$
and $\epsilon^{}_3$, they will be assumed to vanish in the following discussions.
For the sake of simplicity, the hat symbols on $\hat{a}^{}_{1,2}$, $\hat{b}^{}_{1,2}$ and
$\hat{\phi}^{}_{a,b}$ will also be neglected. Consequently, we are led to an effective
neutrino mass matrix $M^\prime_\nu$ of the form
\begin{eqnarray}
M^\prime_{ee} & = & a^2_1 \left( 1 + 2 {\rm i}\epsilon^{}_1 \right) + b^2_1 \;,\nonumber\\
M^\prime_{ e \mu } & = & a^{}_1 a^{}_2  e^{{\rm i} \phi^{}_a} \left( 1+ {\rm i} \epsilon^{}_1
+ \epsilon^{}_3 \right) + b^{}_1 b^{}_2  e^{{\rm i} \phi_b} \;, \nonumber\\
M^\prime_{e\tau} & = & a^{}_1 a^{}_2 e^{ -{\rm i} \phi^{}_a} \left( 1+ {\rm i} \epsilon^{}_1
- \epsilon_3^* \right) + b^{}_1 b^{}_2  e^{ -{\rm i} \phi^{}_b} \;, \nonumber \\
M^\prime_{ \mu \tau } & = &  a^{2}_2 \left[ 1 + 2 {\rm i} {\rm Im}(\epsilon^{}_3)\right]
+ b^{2}_2 \;, \nonumber\\
M^\prime_{ \mu \mu } & = & a^{2}_2 e^{2 {\rm i} \phi^{}_a} \left( 1 + 2\epsilon^{}_3 \right)
+ b^{2}_2 e^{2{\rm i} \phi^{}_b} \;, \nonumber\\
M^\prime_{ \tau \tau } & = & a^{2}_2 e^{-2 {\rm i} \phi^{}_a} \left( 1 - 2\epsilon_3^* \right)
+ b^{2}_2 e^{-2{\rm i} \phi^{}_b} \;,
\label{22}
\end{eqnarray}
at the leading order.
The unitary matrix $U^\prime_\nu$ for diagonalizing $M^\prime_\nu$ is expected to have
some mixing parameters around the special values given by Eq. (\ref{9}),
with the corresponding deviations
\begin{eqnarray}
&& \Delta \phi^{}_1 = \phi^\prime_1 - 0 \;, \hspace {1cm}
\Delta \phi^{}_2 = \left(\phi^\prime_2 + \phi^\prime_3 \right) /2 - 0 \;,
\hspace {1cm} \Delta \theta^{}_{23} = \theta^\prime_{23} - \pi/4 \;,  \nonumber \\
&& \Delta \delta = \delta^\prime - \delta \;, \hspace {1.35cm}
\Delta \sigma = \sigma^\prime - \sigma \;,
\label{23}
\end{eqnarray}
being some small quantities. By making series expansions for these
mixing-parameter deviations in the diagonalization process, at the leading order
we acquire the following relations connecting them with the symmetry-breaking parameters
$\epsilon^{}_{1, 3}$
\begin{eqnarray}
&&  m^{}_3 s^2_{13}  \Delta \delta + \overline m^{}_2 s^2 _{12} \Delta \sigma
= \mp \left[ a^2_1 \epsilon^{}_1 - \left( a^2_1 + b^2_1 \right) \Delta \phi^{}_1 \right] \;,
\nonumber \\
&& \sqrt 2 \left[ \left( m^{}_1 - \overline m^{}_2 \right) c^{}_{12} s^{}_{12}
+ {\rm i} \left( m^{}_1 c^2_{12} + \overline m^{}_2 s^2_{12} + m^{}_3 \right) \bar{s}^{}_{13} \right] \Delta \theta^{}_{23}
\nonumber \\
&& \hspace{0.5cm} - \sqrt{2} \left( m^{}_1 c^2_{12} + \overline m^{}_2 s^2_{12}
- m^{}_3 \right) \bar{s}^{}_{13} \Delta \delta + 2 \sqrt{2} \overline m^{}_2 s^{}_{12}
\left( {\rm i} c^{}_{12} - s^{}_{12} \bar{s}^{}_{13} \right)  \Delta \sigma \; ,
\nonumber \\
&& \hspace{0.5cm} = \mp 2\left[ a^{}_1 a^{}_2 e^{ {\rm i} (\phi^{}_a-\phi) } \left( {\rm i} \epsilon^{}_1
+ \epsilon^{}_3 - {\rm i} \Delta \phi^{}_1 -  {\rm i} \Delta \phi^{}_2  \right)
+ b^{}_1 b^{}_2 e^{ {\rm i} (\phi^{}_b-\phi) } \left(-{\rm i} \Delta \phi^{}_1 - {\rm i}
\Delta \phi^{}_2 \right) \right] \;,
\nonumber \\
&& - \left( m^{}_1 s^2_{12} + \overline m^{}_2 c^2_{12} - m^{}_3 \right) \Delta \theta^{}_{23}
+ \left[ \left( m^{}_1 - \overline m^{}_2 \right) c^{}_{12} s^{}_{12} - {\rm i} \left( m^{}_1 c^2_{12}
+ \overline m^{}_2 s^2_{12} \right) \bar{s}^{}_{13} \right] \bar{s}^{}_{13} \Delta \delta
\nonumber \\
&& \hspace{0.5cm} + \overline m^{}_2 c^{}_{12} \left( {\rm i} c^{}_{12}
- 2 s^{}_{12} \bar{s}^{}_{13} \right) \Delta \sigma = \mp 2
\left[ a^2_2 e^{ {2 {\rm i} (\phi^{}_a-\phi)} } \left( \epsilon^{}_3 - {\rm i} \Delta \phi^{}_2 \right)
+ b^2_2 e^{ {2 {\rm i} (\phi^{}_b-\phi)} } \left( - {\rm i} \Delta \phi^{}_2 \right) \right] \; ,
\label{24}
\end{eqnarray}
where $m^{}_1 = 0$ (or $m^{}_3 = 0$) in the NH (or IH) case and the values of $a^{}_{1, 2}$,
$b^{}_{1, 2}$ and $\phi^{}_{a, b} -\phi$ are the same as those presented in Fig. 1.
The sign $\mp$ which takes $-$ (or $+$) in the NH (or IH) case arises from the
aforementioned fact that $M^{}_\nu$ might need an overall sign change so as to fit in with
the numerical results.

By solving these equations in a straightforward way,
one will obtain the mixing-parameter deviations as some linear functions of
$\epsilon^{}_1$, ${\rm Re}(\epsilon^{}_3)$ and ${\rm Im}(\epsilon^{}_3)$.
For illustration, in Fig. 2 we present the $\Delta \theta^{}_{23}$,
$\Delta \delta$ and $\Delta \sigma$ (as functions of $a^{}_1$) arising from $\epsilon^{}_1 = 0.1$,
${\rm Re}(\epsilon^{}_3) =0.1$ and ${\rm Im}(\epsilon^{}_3) =0.1$ in the NH and IH cases.
Provided that the linear approximation holds to a good degree (i.e., the expected
small quantities are really $\le \mathcal O(0.1)$), the $\Delta \theta^{}_{23}$,
$\Delta \delta$ and $\Delta \sigma$ generated by other values of $\epsilon^{}_1$,
${\rm Re}(\epsilon^{}_3)$ and ${\rm Im}(\epsilon^{}_3)$ can be inferred by rescaling
these results (according to the linear dependence of mixing-parameter deviations
on symmetry-breaking parameters). The results in Fig. 2 tell us:
(a) In the NH case, $\epsilon^{}_1 = 0.1$ may give rise to a $|\Delta \delta|$ as
large as 0.1. But the resulting $|\Delta \theta^{}_{23}|$
and $|\Delta \sigma|$ are desperately small. (b) In the IH case, the $|\Delta \theta^{}_{23}|$,
$|\Delta \delta|$ and $|\Delta \sigma|$ from $\epsilon^{}_1 = 0.1$ are
$\simeq \mathcal O(0.01)$. (c) In the NH case, ${\rm Re}(\epsilon^{}_3) = 0.1$ likely leads to
some considerable ($\simeq 0.1$ or so) $|\Delta \theta^{}_{23}|$, $|\Delta \delta|$ and
$|\Delta \sigma|$. (d) In the IH case, the $|\Delta \theta^{}_{23}|$ and $|\Delta \delta|$
induced by ${\rm Re}(\epsilon^{}_3) = 0.1$ may reach 0.1 and 0.35 (but
for distinct values of $a^{}_1$), while $|\Delta \sigma|$ is
rather small. (e)-(f) In both the NH and IH cases, ${\rm Im}(\epsilon^{}_3) = 0.1$ can result in
considerable $|\Delta \delta|$ and $|\Delta \sigma|$ but relatively small
$|\Delta \theta^{}_{23}|$. In the particular case of $a^{}_1 = 0$,
${\rm Re}(\epsilon^{}_3) = 0.1$ contributes $|\Delta \theta^{}_{23}| \simeq 0.06 \ ({\rm or}\ 0.10)$,
$|\Delta \delta| \simeq 0.21 \ ({\rm or}\ 0.01)$ and $|\Delta \sigma| \simeq 0.14 \ ({\rm or}\ 0.02)$ for NH (or IH), while
${\rm Im}(\epsilon^{}_3) = 0.1$ contributes $|\Delta \theta^{}_{23}| \simeq 0.03 \ ({\rm or}\ 0.00)$,
$|\Delta \delta| \simeq 0.12 \ ({\rm or}\ 0.10)$ and $|\Delta \sigma| \simeq 0.05 \ ({\rm or}\ 0.00)$.
To summarize, $\epsilon^{}_1$ is unlikely to induce considerable
mixing-parameter deviations, while ${\rm Re}(\epsilon^{}_3)$ is
likely. Inversely, a considerable $\Delta \theta^{}_{23}$
can be ascribed to ${\rm Re}(\epsilon^{}_3)$, while a considerable
$\Delta \delta$ may result from any symmetry-breaking parameter.

\begin{figure}
\centering
\includegraphics[width=6in]{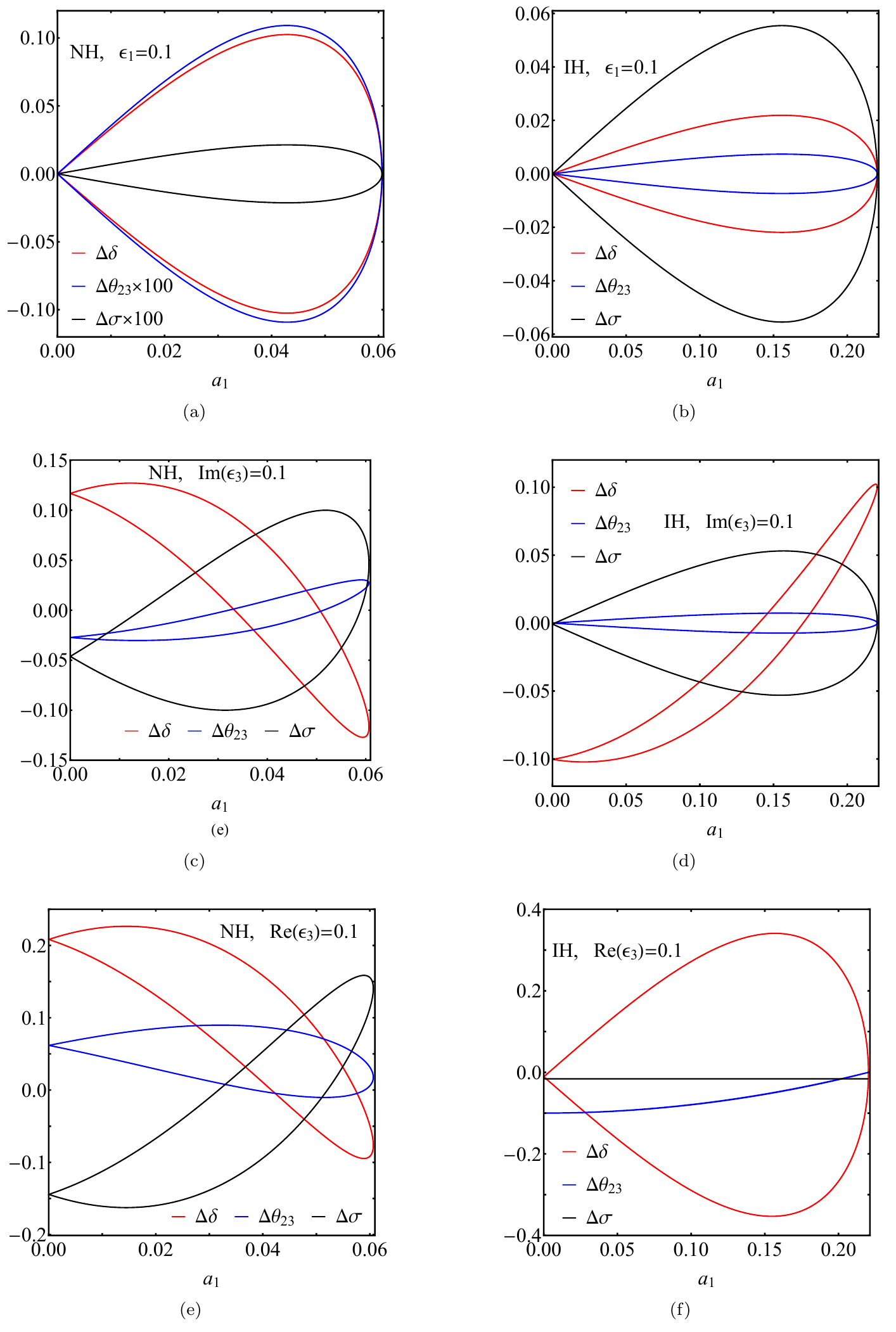}
\caption{ The $\Delta \theta^{}_{23}$, $\Delta \delta $ and $\Delta \sigma$
(as functions of $a^{}_1$) induced by $\epsilon^{}_1 = 0.1$, ${\rm Re}(\epsilon^{}_3) =0.1$
and ${\rm Im}(\epsilon^{}_3) =0.1$ in the NH and IH cases.}
\label{fig:subfig} 
\end{figure}

\section{RGE triggered symmetry breaking}

A flavor symmetry (FS) such as the $\mu$-$\tau$ reflection symmetry under study
is usually introduced at an energy scale $\Lambda^{}_{\rm FS}$ much higher than
the electroweak (EW) one $\Lambda^{}_{{\rm EW}} \sim 10^2$ GeV \cite{review}.
Therefore, the RGE effects should be taken into account when the flavor-symmetry
model is confronted with the low-energy experimental data \cite{OZ}.
During the evolution process, the significant difference between the Yukawa coupling of
$\mu$ and that of $\tau$ may provide a unique source for symmetry breaking.
This section is just devoted to such a specific breaking of the $\mu$-$\tau$
reflection symmetry. At the one-loop level, the energy dependence of $M^{}_{\nu}$
is described by \cite{RGE}
\begin{eqnarray}
16 \pi^2 \frac{ {\rm d} M^{}_{\nu} }{{\rm d} t} = C \left( Y^{\dagger}_{l} Y^{}_{l} \right)^{\rm T} M^{}_{\nu}
+ C M^{}_{\nu} \left( Y^{\dagger}_{l} Y^{}_{l} \right) + \alpha M^{}_{\nu} \; ,
\label{25}
\end{eqnarray}
where $t\equiv {\rm ln} \left( \mu / \mu^{}_0 \right)$ with $\mu$ denoting the
renormalization scale, $C$ and $\alpha$ read
\begin{eqnarray}
&& C= - \frac{3}{2} \;,  \hspace{1cm} \alpha \simeq -3 g^2_2 + 6 y^2_t + \lambda \;,
\hspace{1cm} {\text{in the SM}}  \;; \nonumber \\
&& C= 1 \;, \hspace{1cm} \alpha \simeq -\frac{6}{5} g^2_1 - 6 g^2_2 + 6 y^2_t  \;,
\hspace{1cm} {\text{in the MSSM}} \;.
\label{26}
\end{eqnarray}
In the basis of $M^{}_l$ being diagonal, the Yukawa coupling matrix for charged leptons
is given by $Y^{}_l = {\rm Diag} (y^{}_e, y^{}_\mu, y^{}_\tau)$. Due to
$y^{}_e \ll y^{}_\mu \ll y^{}_\tau$, it is reasonable to neglect the contributions of $y^{}_e$
and $y^{}_\mu$. In Eq. (\ref{25}), the $\alpha$-term is flavor universal and only
contributes an overall rescaling factor $I^{}_{\alpha}$ for the neutrino mass matrix,
while the other two terms are able to modify its structure.
Given an $M^{}_{\nu}(\Lambda^{}_{\rm FS})$ of the form given by Eq. (\ref{14})
at $\Lambda^{}_{\rm FS}$, integration of Eq. (\ref{25}) enables us to derive
the RGE-corrected neutrino mass matrix at $\Lambda^{}_{{\rm EW}}$ \cite{IRGE}
\begin{eqnarray}
M^{}_{\nu} (\Lambda^{}_{\rm EW}) & = & I^{}_{\alpha} I^{\dagger}_{\tau}
M^{}_{\nu} (\Lambda^{}_{\rm FS}) I^{*}_{\tau} \nonumber \\
& = & I^{}_{\alpha} \left[ M^{}_{\nu} ( \Lambda^{}_{\rm FS} ) - \Delta^{}_{\tau}
\left( \begin{matrix}
0 & 0 & M^{}_{ e \tau} \cr
0 & 0 & M^{}_{ \mu \tau} \cr
M^{}_{ e \tau} & M^{}_{ \mu \tau} & 2 M^{}_{ \tau \tau}  \cr
\end{matrix} \right) \right] \;,
\label{27}
\end{eqnarray}
with $I^{}_{\tau}\simeq {\rm Diag} (1,1, 1-\Delta^{}_{\tau} )$ and
\begin{eqnarray}
I^{}_{\alpha} = {\rm exp} \left( - \frac{1}{16 \pi^2} \int_{\rm ln \  \Lambda^{}_{EW}}^{\rm ln
\ \Lambda^{}_{\rm FS}} \alpha \ {\rm dt} \right) \;, \hspace{1cm}
\Delta^{}_{\tau} = \frac{C}{16 \pi^2}\int_{\rm ln \  \Lambda^{}_{\rm EW} }^{\rm ln
\ \Lambda^{}_{FS} } y^2_{\tau} \ {\rm dt} \;.
\label{28}
\end{eqnarray}
Obviously, $\Delta^{}_{\tau}$ measures the strength of symmetry breaking.
Owing to the smallness of $y^{}_\tau \simeq 0.01$ which gives
$\Delta^{}_\tau \simeq \mathcal O(10^{-5})$,
the RGE effect is negligible in the SM. But in the MSSM,
$y^{2}_\tau = (1+ \tan^2{\beta}) m^2_\tau/v^2$ (with $v = 174$ GeV being the Higgs VEV)
can be greatly enhanced by a large $\tan{\beta}$.
To be explicit, the value of $\Delta^{}_{\tau}$ depends on $\tan{\beta}$ in a way as
\begin{eqnarray}
\Delta^{}_{\tau} \simeq 0.042 \left( \displaystyle \frac{\tan{\beta}}{50} \right)^2 \;,
\label{29}
\end{eqnarray}
if we take $\Lambda^{}_{\rm FS} \simeq 10^{13}$ GeV as an example.

Following the same approach as in the previous section, one can obtain the following relations
connecting the mixing-parameter deviations with $\Delta^{}_\tau$
\begin{eqnarray}
&& m^{}_3 s^2_{13} \Delta \delta + \overline m^{}_2 s^2 _{12} \Delta \sigma
= \left( m^{}_1 c^2_{12} + \overline m^{}_2 s^2_{12} - m^{}_3 s^2_{13} \right) (-\Delta \phi^{}_1)    \; ,
\nonumber \\
&& 2 \left[ \left( m^{}_1 - \overline m^{}_2 \right) c^{}_{12} s^{}_{12}
+ {\rm i} \left( m^{}_1 c^2_{12} + \overline m^{}_2 s^2_{12}
+ m^{}_3 \right) \bar{s}^{}_{13} \right] \Delta \theta^{}_{23}
\nonumber \\
&& \hspace{1cm} - 2\left( m^{}_1 c^2_{12} + \overline m^{}_2 s^2_{12}
- m^{}_3 \right) \bar{s}^{}_{13} \Delta \delta + 4 \overline m^{}_2 s^{}_{12}
\left( {\rm i} c^{}_{12} - s^{}_{12} \bar{s}^{}_{13} \right)  \Delta \sigma \; ,
\nonumber \\
&& \hspace{1cm} = \left[ {\rm i} (m^{}_{11} + m^{}_3) \bar s^{}_{13} - m^{}_{12} \right]
(\Delta^{}_\tau - 2{\rm i} \Delta \phi^{}_1 - 2{\rm i} \Delta \phi^{}_2) \;,
\nonumber \\
&& - 2 \left( m^{}_1 s^2_{12} + \overline m^{}_2 c^2_{12} - m^{}_3 \right) \Delta \theta^{}_{23}
+ 2 \overline m^{}_2 c^{}_{12} \left( {\rm i} c^{}_{12} - 2 s^{}_{12} \bar{s}^{}_{13} \right) \Delta \sigma
\nonumber \\
&& \hspace{1cm} + 2 \left[\left( m^{}_1 - \overline m^{}_2 \right) c^{}_{12} s^{}_{12}
- {\rm i} \left( m^{}_1 c^2_{12} + \overline m^{}_2 s^2_{12} \right)
\bar{s}^{}_{13} \right] \bar{s}^{}_{13} \Delta \delta
\nonumber \\
&& \hspace{1cm} = [ m^{}_1 s^2_{12} + \overline m^{}_2 c^2_{12} + m^{}_3 -
2{\rm i} (m^{}_1 - \overline m^{}_2) c^{}_{12} s^{}_{12} \bar s^{}_{13} ]
(\Delta^{}_\tau - 2{\rm i}\Delta \phi^{}_2) \;,
\label{30}
\end{eqnarray}
with $m^{}_1 = 0$ (or $m^{}_3 = 0$) in the NH (or IH) case. Solving these equations
gives
\begin{eqnarray}
&& \Delta \theta^{}_{23} = + 0.40 \Delta^{}_{\tau} \;, \hspace{1cm} \Delta \delta = + 0.66 \Delta^{}_{\tau} \;,
\hspace{1cm} \Delta \sigma = -0.02  \Delta^{}_{\tau} \;,
\end{eqnarray}
in the NH case, or
\begin{eqnarray}
&& \Delta \theta^{}_{23} = - 0.52 \Delta^{}_{\tau} \;, \hspace{1cm} \Delta \delta = -0.06 \Delta^{}_{\tau} \;,
\hspace{1cm} \Delta \sigma = -0.16 \Delta^{}_{\tau} \;,
\end{eqnarray}
in the IH case. We subsequently show the dependence of these mixing-parameter deviations
on the value of $\tan{\beta}$ (which varies from 10 to 50) in Fig. 3. One can see that
the mixing parameters are pretty stable against the RGE corrections.
Even for $\tan{\beta} = 50$, one merely has $|\Delta \theta^{}_{23}| \simeq 0.017 \ ({\rm or} \ 0.021)$,
$|\Delta \delta| \simeq 0.028 \ ({\rm or} \ 0.003)$ and
$|\Delta \sigma| \simeq 0.001 \ ({\rm or} \ 0.007)$ in the NH (or IH) case.

\begin{figure}
\centering
\includegraphics[width=6in]{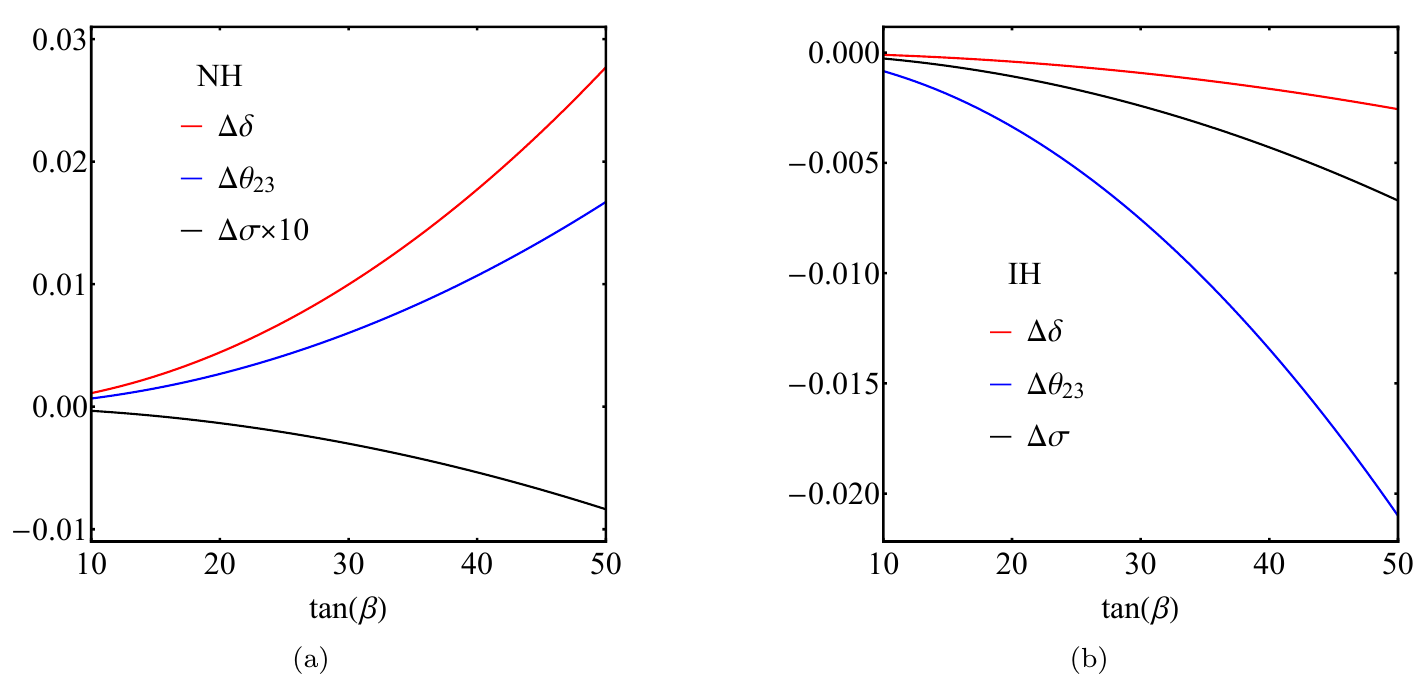}
\caption{ The RGE-induced $\Delta \theta^{}_{23}$, $\Delta \delta$
and $\Delta \sigma$ against $\tan \beta$ in the MSSM.}
\label{fig:subfig} 
\end{figure}

\section{Leptogenesis and the symmetry breaking}

Besides accommodating the smallness of neutrino masses, the seesaw mechanism
can also explain the observed baryon-antibaryon asymmetry of the Universe via the
leptogenesis mechanism \cite{leptogenesis}: The CP-violating, lepton-number-violating
and out-of-equilibrium decays of $N^{}_i$ may generate a
lepton-antilepton asymmetry which is eventually converted to the baryon-antibaryon
asymmetry through the sphaleron process \cite{Yanagida}. The essential CP violation
is provided by the complex Yukawa couplings $Y^{}_{\rm D} = M^{}_{\rm D}/v$
which couple $N^{}_i$ with the left-handed neutrino fields, while the lepton-number
violation originates from the Majorana mass terms of $N^{}_i$. And
the departure from thermal equilibrium can occur if $N^{}_i$
decay in a rate smaller than the expansion rate of the Universe when the
temperature $T$ drops to the mass scale of $N^{}_i$. The produced amount of lepton-antilepton
asymmetry is crucially dependent on the CP-violating asymmetries
between the decays of $N^{}_i$ and their CP conjugate processes.
In the minimal seesaw under study, we assume $N^{}_{1, 2}$ to have
a strong mass hierarchy $M^{}_1 \ll M^{}_2$, in which case only the decay of $N^{}_1$
is relevant for leptogenesis. The flavored CP-violating asymmetries
$\varepsilon_{\alpha}$ are given by \cite{CP-asymmetry}
\begin{equation}
\varepsilon_{\alpha} \simeq - \frac{3}{16 \pi ( Y_{\rm D}^{\dagger} Y^{}_{\rm D} )_{11} }
{ \rm Im} \left[ ( Y_{\rm D}^{\dagger} Y^{}_{\rm D} )_{12} \  ( Y_{\rm D}^{\dagger} )_{1 \alpha}
(Y^{}_{\rm D})_{ \alpha 2} \right] \frac{ M^{}_1 }{ M^{}_2 } \;,
\end{equation}
for $\alpha = e, \mu, \tau$. A $Y^{}_{\rm D}$ corresponding to the $M^{}_{\rm D}$
in Eq. (\ref{13}) immediately yields \cite{MTR}
\begin{equation}
\varepsilon^{}_e = 0 \;, \hspace{1cm}  \varepsilon^{}_{\mu} = - \varepsilon^{}_{\tau} \;,
\end{equation}
rendering the total CP-violating asymmetry
$\varepsilon = \varepsilon^{}_{e} + \varepsilon^{}_{\mu} + \varepsilon^{}_{\tau}$ vanishing.
However, a successful leptogenesis is possible
when the $\mu$-$\tau$ reflection symmetry is broken \cite{ahn} or (and) the flavor effects
become relevant \cite{mohapatra2}.

For the situation of $M^{}_1 > 10^{12}$ GeV, the flavor effects are negligible.
In order to achieve $\varepsilon \neq 0$, one has to break the $\mu$-$\tau$ reflection symmetry.
A $Y^{\prime}_{\rm D}$ corresponding to the $M^{\prime}_{\rm D}$
in Eq. (\ref{19}) (with $\epsilon^{}_{2, 4}=0$ as assumed before) gives
\begin{eqnarray}
\varepsilon \simeq && \frac{3 M^{}_1 \left[ a^{}_1 b^{}_1
+ 2 a^{}_2 b^{}_2 \cos \left( \phi^{}_a - \phi^{}_b \right) \right] }
{ 8 v^2 \left( a^2_1 + 2 a^2_2 \right) }
\nonumber  \\
&& \times \left[ a^{}_1 b^{}_1 \epsilon^{}_1 + 2 a^{}_2 b^{}_2
\sin \left( \phi^{}_a - \phi^{}_b \right) {\rm Re} \left( \epsilon^{}_3 \right)
+ 2 a^{}_2 b^{}_2 \cos \left( \phi^{}_a - \phi^{}_b \right)
{\rm Im} \left( \epsilon^{}_3 \right) \right] \;,
\end{eqnarray}
which means that $\varepsilon$ is proportional to $M^{}_1$ and a linear function of
$ \epsilon^{}_1$, $ {\rm Re} \left( \epsilon^{}_3 \right) $ and
${\rm Im} \left( \epsilon^{}_3 \right)$.
The final baryon-to-entropy ratio can be written as \cite{Yanagida}
\begin{equation}
{\rm Y^{}_{ B} } \equiv \frac{ n^{}_{\rm B} - n^{}_{\rm \bar B} } {s}
\simeq - \frac{12}{37} \kappa \frac{\varepsilon}{ g^{}_{*} } \;.
\end{equation}
Here $12/37$ is the efficiency factor of converting the lepton-antilepton asymmetry
to the baryon-antibaryon asymmetry, whereas $g_{*}=106.75$ is the effective number of
relativistic degrees of freedom at $T=M^{}_1$ in the SM. In particular, $\kappa$ is the
washout factor and can be parametrized as \cite{Yanagida}
\begin{equation}
\kappa \simeq \left( 2 \pm 1 \right) \times 10^{-2} \times
\left( \frac{{\rm 0.01 \ eV}}{\tilde{m}^{}_1}\right)^{ 1.1 \pm 0.1} \;,
\end{equation}
with $ \tilde{m}^{}_1 = ( Y^{\dagger}_{\rm D} Y^{}_{\rm D} )^{}_{11} v^2 / M^{}_1 $.
In the present epoch (for $s = 7.04 n^{}_\gamma$), the baryon-to-photon ratio
is given by
\begin{equation}
\eta \equiv \frac{ n^{}_{\rm B} - n^{}_{\rm \bar{B}} }
{ n^{}_{\gamma} } \simeq  7.04 {\rm Y^{}_B} \;,
\end{equation}
which has an observed value of $(6.08 \pm 0.09) \times 10^{-10}$ \cite{WMAP}.
To figure out what kind of $M^{}_1$ and symmetry-breaking parameters
may give rise to the observed baryon-antibaryon asymmetry, in Fig. 4 we present the $\eta$
(as functions of $a^{}_1$) arising from some example values of them.
The results show that a combination of
$|M^{}_1 \epsilon^{}_1| \simeq \mathcal O(10^{11}) \ ({\rm or} \ \mathcal O(10^{13}))$ GeV,
$|M^{}_1 {\rm Im}(\epsilon^{}_3)| \simeq \mathcal O(10^{10}) \ ({\rm or} \ \mathcal O(10^{13}))$ GeV or
$|M^{}_1 {\rm Re}(\epsilon^{}_3)| \simeq \mathcal O(10^{10}) \ ({\rm or} \ \mathcal O(10^{13}))$ GeV
can give successful leptogenesis in the NH (or IH) case. Clearly, it is much easier
to gain the observed $\eta$ in the NH case than in the IH case.
In particular, a $|{\rm Re}(\epsilon^{}_3)| \ {\rm or} \
|{\rm Im}(\epsilon^{}_3)| \simeq \mathcal O(0.01)$ (or smaller
if $M^{}_1$ takes a value larger than $10^{12}$ GeV)
is sufficient for generating the observed $\eta$ in the NH case.

\begin{figure}
\centering
\includegraphics[width=6in]{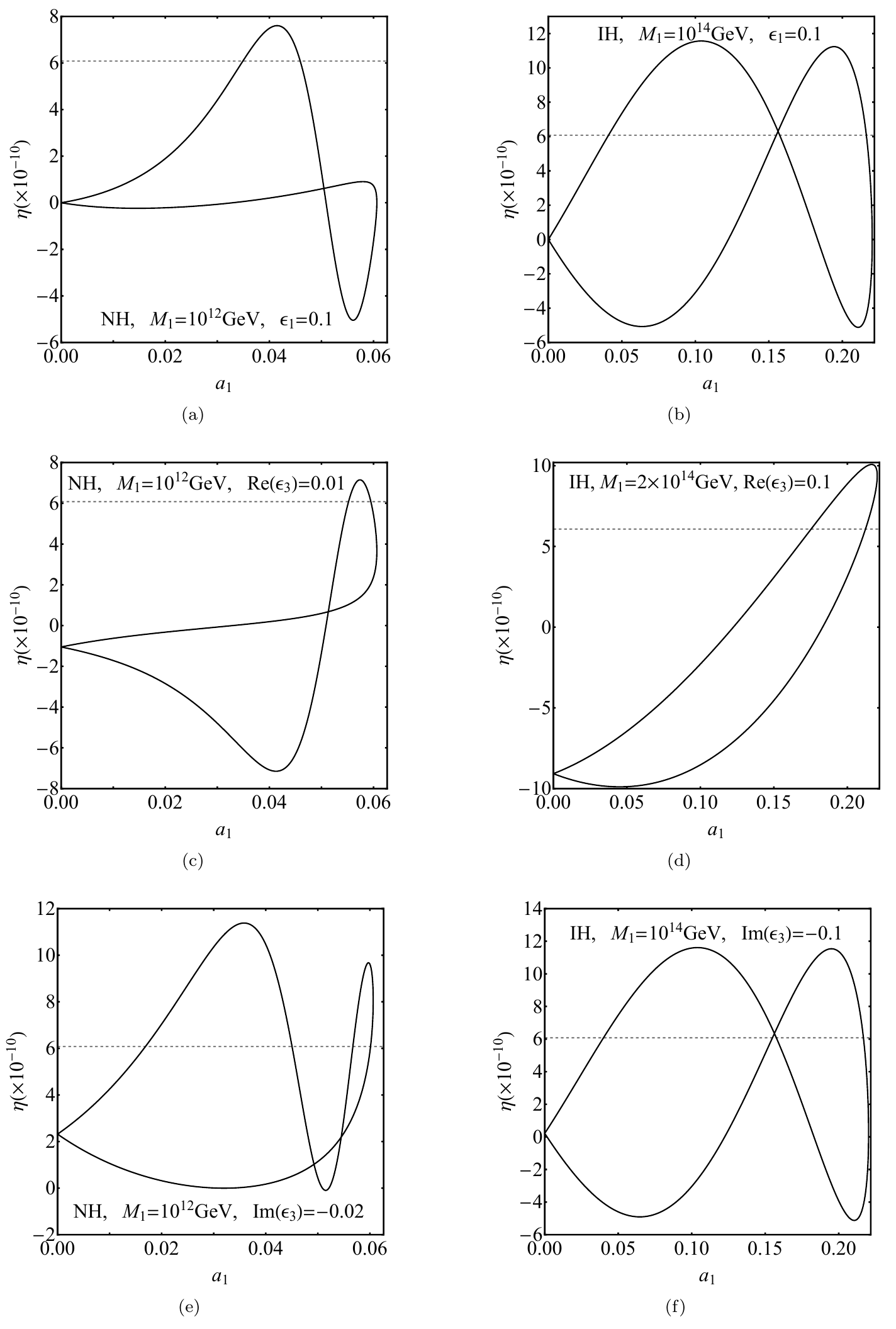}
\caption{The $\eta$ (as functions of $a^{}_1$) arising from some example
values of $M^{}_1$ and symmetry-breaking parameters for the situation of
$M^{}_1 > 10^{12}$ GeV. The gray dashed line stands for the observed
value of $\eta$. }
\label{fig:subfig} 
\end{figure}

If $M^{}_1$ turns out to be smaller than $10^{12}$ GeV,
then the Yukawa interactions of charged leptons will enter in thermal equilibrium,
making different lepton flavors distinguishable.
In such a case, both the CP-violating asymmetries and
washout factors associated with different flavors should be treated separately.
Accordingly, the final baryon-to-entropy ratio can be rewritten as \cite{flavor-effects}
\begin{eqnarray}
{\rm Y_B} \simeq - \frac{12}{37} \frac{K}{g^{}_*} \;,
\end{eqnarray}
where the parameter $K$ is defined as
\begin{eqnarray}
&&  K = \left( \epsilon^{}_e + \epsilon^{}_{\mu} \right) \kappa^{}_f \left( \frac {417}{589}
\tilde{m}^{}_e + \frac {417}{589} \tilde{ m}^{}_{\mu}  \right) + \epsilon^{}_{\tau} \kappa^{}_f
\left( \frac{390}{589} \tilde{m}^{}_{\tau} \right) \;, \hspace{0.2cm}
{\rm if} \hspace{0.2cm} 10^{9} \ {\rm GeV}  <  M^{}_1 < 10^{12} \ {\rm GeV} \;;
\nonumber \\
&& K = \epsilon^{}_e \kappa^{}_f \left( \frac {151}{179} \tilde{m}^{}_e \right)
+ \epsilon^{}_{\mu} \kappa^{}_f \left( \frac{344}{537} \tilde{m}^{}_{\mu} \right)
+ \epsilon^{}_{\tau} \kappa^{}_f \left( \frac{344}{537} \tilde{m}^{}_{\tau} \right) \;,
\hspace{0.2cm} {\rm if} \hspace{0.2cm} M^{}_1 < 10^9 \ {\rm GeV} \;,
\end{eqnarray}
with
\begin{eqnarray}
\kappa^{}_f \left( \tilde{m}^{}_{\alpha} \right) \simeq
\left[ \left( \frac{ \tilde{m}^{}_{\alpha} }{ 2.1 \times 10^{-3} \ {\rm eV} } \right)^{-1}
+ \left( \frac{5 \times 10^{-4} \ {\rm eV}}{ \tilde{m}^{}_{\alpha}} \right)^{-1.16}
\right]^{-1} \;, \hspace{0.2cm} {\rm for} \hspace{0.2cm} \tilde{m}^{}_{\alpha} \equiv
\frac { | (Y^{}_{\rm D})^{}_{\alpha 1} |^2 v^2 }{M^{}_1} \;.
\end{eqnarray}
For illustration, in Fig. 5 we show the $\eta$ (as functions of $a^{}_1$) arising from
some example values of $M^{}_1$ and symmetry-breaking parameters for the situation
of $10^9 \ {\rm GeV} < M^{}_1 < 10^{12} \ {\rm GeV}$.
For comparison, the contributions from pure flavor effects
(without symmetry-breaking effects) are also shown.
It is easy to see that the flavor effects are more significant than
the symmetry-breaking effects in the NH case, while the contrary
is the case in the IH case. In the NH case the flavor effects themselves are competent
for generating the observed $\eta$. But in the IH case the symmetry-breaking
effects have to be invoked and the symmetry-breaking parameters should take some
values at least $\mathcal O(0.1)$.

\begin{figure}
\centering
\includegraphics[width=6in]{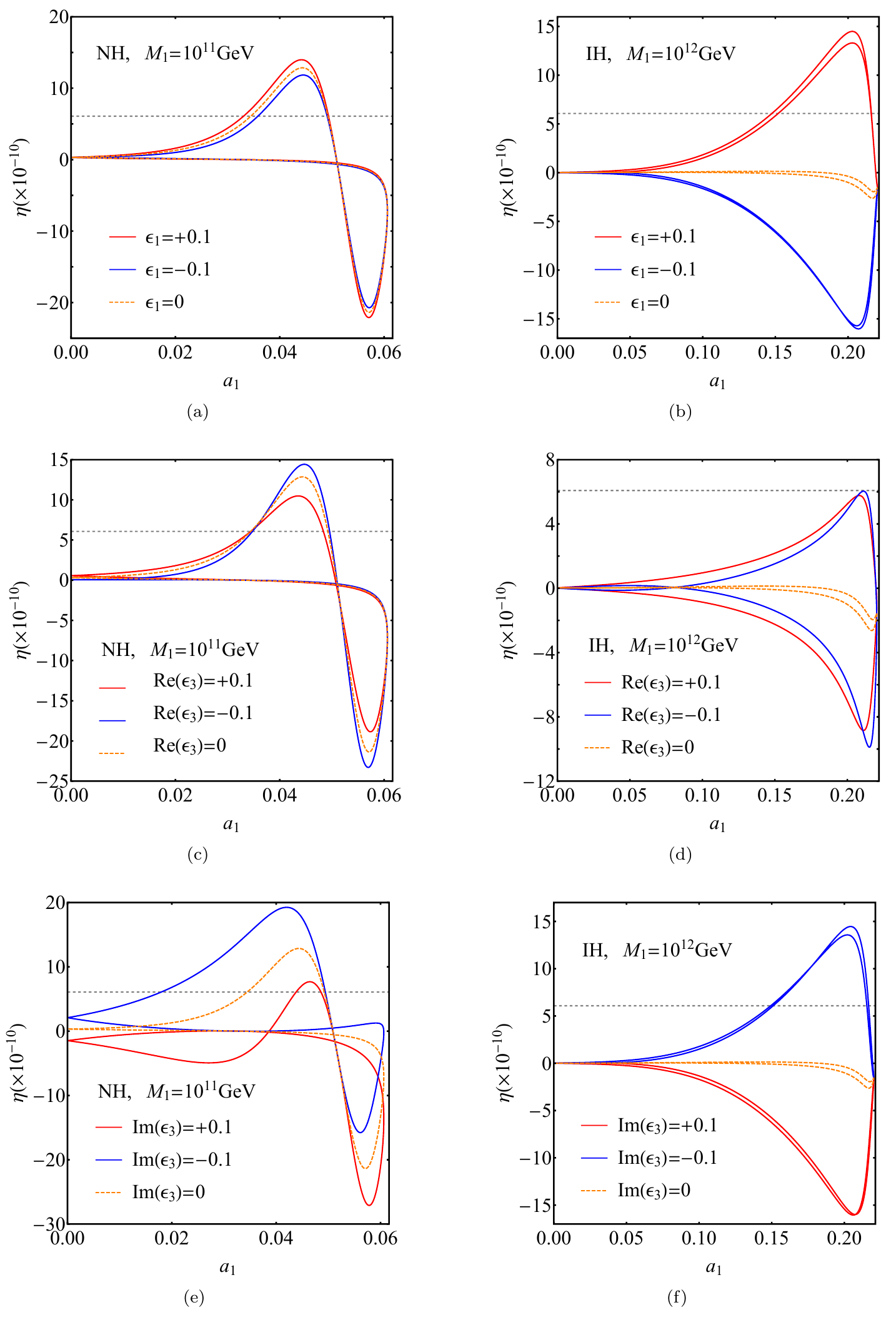}
\caption{The $\eta$ (as functions of $a^{}_1$) arising from some example
values of $M^{}_1$ and symmetry-breaking parameters for the situation of
$10^9 \ {\rm GeV < M^{}_1 < 10^{12} \ {\rm GeV} }$.
The gray dashed line stands for the observed value of $\eta$. }
\label{fig:subfig} 
\end{figure}

In order to see whether there exists a choice of basic
parameters which can lead to sizable $\Delta \delta$,
$\Delta \theta^{}_{23}$ and successful leptogenesis, in Figs. 6-7 we present
the resulting $\Delta \delta$ and $\Delta \theta^{}_{23}$ from
the parameter choices shown in Figs. 4-5 that can lead to successful leptogenesis.
From Fig. 6 one finds that for the situation of $M^{}_1 > 10^{12}$ GeV
a sizable $\Delta \delta$ can be generated in association with a realistic $\eta$ from
$\epsilon^{}_1$ in the NH case (see the sub-figure labelled as (a)),
${\rm Re}(\epsilon^{}_3)$ in the IH case (see the sub-figure labelled as (d)) or
${\rm Im}(\epsilon^{}_3)$ in the IH case (see the sub-figure labelled as (f)),
but a sizable $\Delta \theta^{}_{23}$ has no chance to arise along with a
successful leptogenesis. The results in Fig. 7 tell us that for the situation of
$10^9 \ {\rm GeV < M^{}_1 < 10^{12} \ {\rm GeV} }$ all the parameter choices except
for that shown in the sub-figure labelled as (b) may give rise to
a sizable $\Delta \delta$, while only the parameter choice shown in
the sub-figure labelled as (c) is capable of producing a sizable $\Delta \theta^{}_{23}$.

\begin{figure}
\centering
\includegraphics[width=6in]{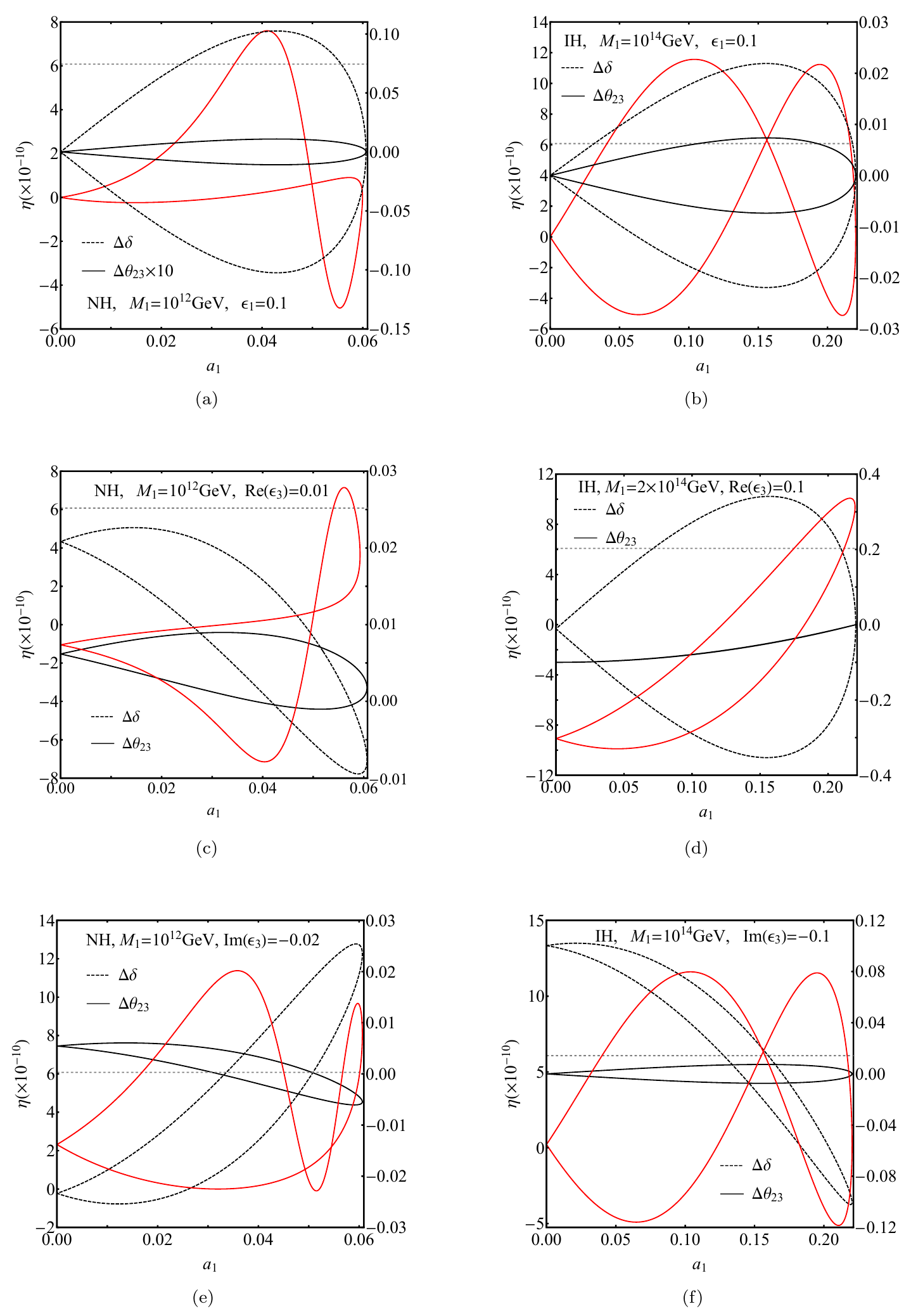}
\caption{The resulting $\Delta \delta$ and $\Delta \theta^{}_{23}$
(as functions of $a^{}_1$) from the parameter choices shown in Fig. 4
that can lead to successful leptogenesis.
The red line is used to denote the corresponding $\eta$. }
\end{figure}

\begin{figure}
\centering
\includegraphics[width=6in]{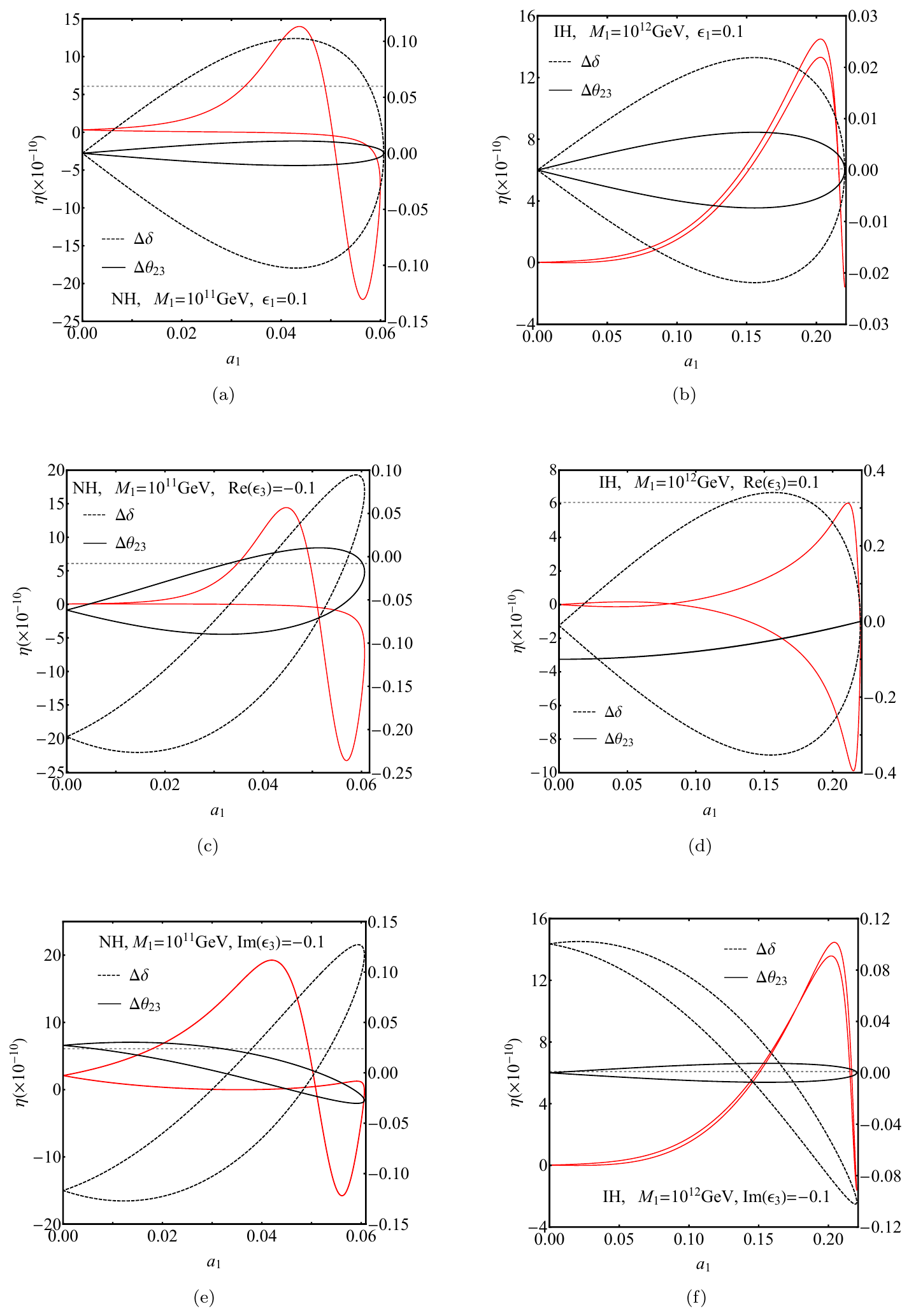}
\caption{The resulting $\Delta \delta$ and $\Delta \theta^{}_{23}$
(as functions of $a^{}_1$) from the parameter choices
shown in Fig. 5 that can lead to successful leptogenesis.
The red line is used to denote the corresponding $\eta$.}
\end{figure}

\section{Summary}

The $\mu$-$\tau$ reflection symmetry is powerful in shaping the neutrino mixing:
it leads to the interesting results $\theta_{23}=\pi/4$ and $\delta= \pm \pi/2$
(which are close to the current experimental results) as well as trivial Majorana phases.
On the other hand, the minimal seesaw has strong predictive power in settling
the neutrino mass spectrum: it enforces the condition of $m^{}_1 =0$ or $m^{}_3 =0$.
In this paper, an attempt of implementing the $\mu$-$\tau$ reflection symmetry
in the minimal seesaw has been made. Such a marriage results in
$\sigma = \pi/2$ (or 0) and thus $|M^{}_{ee}| \simeq 0.37$ (or 4.86) eV
in the NH (or IH) case. Motivated by the preliminary experimental hints towards
$\theta^{}_{23} \neq \pi/4$ and $\delta \neq -\pi/2$, we particularly explore
the possible symmetry breakings and their phenomenological consequences.
Some parameters characterizing the breaking of $\mu$-$\tau$ refection symmetry
are first defined and their implications for the mixing-parameter deviations
then analyzed. It is found that $\epsilon^{}_1$ is difficult to produce
considerable mixing-parameter deviations, while ${\rm Re}(\epsilon^{}_3)$
is relatively easy. Inversely, a considerable $\Delta \theta^{}_{23}$
can be attributed to ${\rm Re}(\epsilon^{}_3)$, while a considerable
$\Delta \delta$ may arise from any symmetry-breaking parameter.

As a unique example, the symmetry breaking triggered by the RGE effects is
studied. It turns out that the mixing parameters are rather
stable against the RGE corrections. Even for $\tan{\beta} \simeq 50$ in the
MSSM, the RGE-induced mixing-parameter deviations are only of $\mathcal O(0.01)$.
Finally, the operation of leptogenesis in the framework under study
is discussed. For the situation of $M^{}_1 > 10^{12}$ GeV where one has a vanishing
$\varepsilon$, the $\mu$-$\tau$ reflection symmetry must be broken to make the
leptogenesis mechanism work. For illustration, we give some example
values of $M^{}_1$ and symmetry-breaking parameters that may give rise to the
observed value of $\eta$.
For the situation of $10^{9} \ {\rm GeV} < M^{}_1 < 10^{12}$ GeV, it is possible that
the flavor effects themselves (without symmetry-breaking effects) are sufficient
for producing the observed $\eta$ in the NH case.
But in the IH case, one has to turn to the symmetry-breaking effects for help.
For both situations, it is easier to achieve a realistic value of $\eta$ in the
NH case than in the IH case. Last but not least, we point out that
the mixing-parameter deviations can be connected to the implementation of leptogenesis,
considering that they may originate from the same symmetry breaking.
The results show that a sizable $\Delta \delta$ can be generated in association
with the observed $\eta$ in many cases, but a sizable $\Delta \theta^{}_{23}$
can only arise along with a successful leptogenesis from ${\rm Re}(\epsilon^{}_3)$
in the NH case for the situation of $10^9 \ {\rm GeV < M^{}_1 < 10^{12} \ {\rm GeV} }$.

\vspace{0.5cm}

\underline{Acknowledgments} \hspace{0.2cm} This work is supported
in part by the National Natural Science Foundation of China under grant
No. 11275088 (Z. C. L and C. X. Y) and grant No. 11605081 (Z. H. Z).

\end{document}